\documentclass[twocolumn,showpacs]{revtex4}
\usepackage{times,xspace}
\usepackage{amsbsy,amssymb,amsmath,bm}
\usepackage{graphicx,color,epsfig,rotate}
\usepackage{fancyheadings}
\def\one{{\mathchoice {\rm 1\mskip-4mu l} {\rm 1\mskip-4mu l} {\rm
1\mskip-4.5mu l} {\rm 1\mskip-5mu l}}}
\def\bbbc{{\mathchoice {\setbox0=\hbox{$\displaystyle\rm C$}\hbox{\hbox
to0pt{\kern0.4\wd0\vrule height0.9\ht0\hss}\box0}}
{\setbox0=\hbox{$\textstyle\rm C$}\hbox{\hbox
to0pt{\kern0.4\wd0\vrule height0.9\ht0\hss}\box0}}
{\setbox0=\hbox{$\scriptstyle\rm C$}\hbox{\hbox
to0pt{\kern0.4\wd0\vrule height0.9\ht0\hss}\box0}}
{\setbox0=\hbox{$\scriptscriptstyle\rm C$}\hbox{\hbox
to0pt{\kern0.4\wd0\vrule height0.9\ht0\hss}\box0}}}}

\newcommand{\Zero}{\ensuremath{\mathbb{O}}\xspace}
\newcommand{\Punitary}{\ensuremath{\mathbb{I}}\xspace}

\newcommand{\ignore}[1]{}
\newcommand{\cComment}[1]{}
\newcommand{\gComment}[1]{}
\renewcommand{\cComment}[1]{\textcolor{blue}{Cristian: #1}}
\renewcommand{\gComment}[1]{\textcolor{red}{Gerardo: #1}}
\begin{document}
\title{Hierarchical Mean-Field Theories in Quantum Statistical
Mechanics}
\author{G. Ortiz and C.D. Batista}
\affiliation{Theoretical Division, 
Los Alamos National Laboratory, Los Alamos, NM 87545}
\date{Received \today }

\begin{abstract}
We present a theoretical framework and a calculational scheme to study
the coexistence and competition of thermodynamic phases in quantum 
statistical mechanics. The crux of the method is the realization that 
the microscopic Hamiltonian, modeling the system, can always be written
in a hierarchical operator language that unveils all symmetry
generators of the problem and, thus, possible thermodynamic phases. In
general one cannot compute the thermodynamic or zero-temperature
properties exactly and an approximate scheme named ``hierarchical
mean-field approach'' is introduced. This approach treats all possible
competing orders on an equal footing. We illustrate the methodology by
determining the phase diagram and quantum critical point of a bosonic
lattice model which displays coexistence and competition between
antiferromagnetism and superfluidity. 
\end{abstract}


\maketitle

During last decade we have witnessed great advances in materials
synthesis and fabrication. The rate at which new compounds with
multiplicity of distinct phases and characteristic functionalities are
generated has outpaced our complete physical understanding of the
fundamental principles behind such complex phenomena. For example, 
whether the mechanism controlling the coexistence and/or competition
between magnetism and superconductivity or Bose-Einstein condensation
has the same physical origin in different classes of materials is still
an open question \cite{mst-10}. On the other hand, the possibility of
control and tunability of the interactions of the elementary
constituents (i.e, quantum control) offers the potential to design new
states of matter with unforeseen applications \cite{greiner}. Despite
great theoretical advances there is a lack of a systematic and reliable
methodology to study and predict the behavior of these complex systems.
It is the main purpose of this paper to present a promising step in
that direction. 

The modern theory of phase transitions starts with Landau's pioneering
work in 1937 \cite{landau}. One of his achievements was the realization
of the fundamental relation between spontaneous symmetry breaking and
the order parameter (OP) that measures this violation, thus giving
simple prescriptions to describe order in terms of irreducible
representations of the symmetry group involved. Another was the
development of a phenomenological calculational scheme to study the
behavior of systems near a phase transition. Landau's
theory has been successfully applied to study phase transitions where
thermal fluctuations are most relevant. Certainly, the theory was not
designed to study zero-temperature (quantum) phase transitions, i.e.,
the qualitative changes of the macroscopic state of the system
induced by tuning parameters of its Hamiltonian. 

In the quantum description of matter, a physical system is naturally
associated with a {\it language} of operators \cite{Not0}. In previous
work \cite{unveiling,rpmbt} we outlined a framework to identify OPs
based upon isomorphic mappings to a {\it hierarchical language} (HL)
defined by the set of operators which in the fundamental representation
(of dimension $D$) has the largest number of symmetry generators of the
group. {\it Any} local operator can be expressed as a {\it linear}
combination of the generators of the HL. The building of the HL
depends upon the dimension $D$ of the local Hilbert space modeling the
physical phenomena one is investigating. For instance, if one is
modeling a doped antiferromagnetic (AF) insulator with a $t$-$J$
Hamiltonian \cite{auerbach}, then $D$=3 (i.e., there are three possible
states per site) and a HL is generated by a basis of $su(3)$ in the
fundamental representation \cite{unveiling,rpmbt}. As explained and
proved in Refs. \cite{nos,rpmbt}, there is always a HL associated to
each physical problem. These ideas complement Landau's concept of an
OP  providing a mechanism to reveal them, something that is outside the
groundwork of Landau's theory. Indeed, this theory does not say what
the OPs should be in a general situation.

It turns out that these isomorphic mappings not only unveil hidden
symmetries of the original physical system but also manifestly
establish equivalences between seemingly unrelated physical phenomena.
Nonetheless, this is not sufficient to determine the {\it exact} phase
diagram of the problem: One has to resort to either numerical
simulations with their well-known limitations or, as will be shown in
the present paper, to a {\it guided} approximation which at least
preserves the qualitative nature of the possible thermodynamic states.
A key observation in this regard is the fact that typical model
Hamiltonian operators written in the HL become quadratic in the
symmetry generators of the hierarchical group, and this result is
independent of the group of symmetries of the Hamiltonian. 

This latter result suggests a simple approximation, based upon group
theoretical grounds, which deals with competing orders on an equal
footing and will be termed {\it hierarchical mean-field theory} (HMFT).
In a sense, that will become clear below, HMFT constitutes the {\it
optimum} mean-field (MF) or saddle-point solution that approximates the
energy and correlation functions of the original problem. The HMFT is
distinctly suitable when the various phases displayed by a system are
the result of competing interactions and non-linear couplings of their
constituents matter fields. From the theoretical standpoint these
systems are strongly correlated since no obvious small coupling
constant exists, as a consequence they exhibit high sensitivity to
small parameter changes. It is then clear the importance of developing
a methodology that treats all possible competing orders on an equal
footing.

We will now illustrate the methodology by example and determine the
zero temperature phase diagram of a simple model which displays
coexistence and competition between antiferromagnetism and
Bose-Einstein condensation (superfluidity). The model represents a gas
of interacting two-flavor ($\sigma= \uparrow,\downarrow$) hard-core
bosons with Hamiltonian ($t>0$)
\begin{eqnarray}
H&=&t \! \sum_{\langle {\bf i},{\bf j} \rangle,\sigma} \left (
{\bar{b}}^\dagger_{{\bf i} \sigma} {\bar{b}}^{\;}_{{\bf j} \sigma} +
{\rm H.c.} \right ) + J \sum_{\langle {\bf i},{\bf j} \rangle} ({\bf
s}_{\bf i} \cdot  {\bf s}_{\bf j} -  \frac{\bar{n}_{\bf i}
{\bar{n}}_{\bf j}}{4})\nonumber \\ &+& V \sum_{\langle {\bf i},{\bf j}
\rangle}  {\bar{n}}_{\bf i} {\bar{n}}_{\bf j}  - \bar{\mu} \sum_{{\bf
j}} {\bar{n}}_{\bf j} \ ,
\label{hamilt}
\end{eqnarray}
where $\langle {\bf i},{\bf j} \rangle$ stands for nearest-neighbor
sites (bond) in an otherwise regular $N_s$-sites lattice of
coordination ${\sf z}$ and $D=3$. The number operator ${\bar {n}}_{\bf
j}=\bar{n}_{{\bf j} \uparrow} + \bar{n}_{{\bf j} \downarrow}$
($\bar{n}_{{\bf j} \sigma}={\bar {b}}^\dagger_{{\bf j} \sigma} {\bar
{b}}^{\;}_{{\bf j} \sigma}$), and ${\bf s}_{\bf j}=\frac{1}{2} {\bar
{b}}^\dagger_{{\bf j} \mu} \boldsymbol{\sigma}_{\mu \nu} {\bar
{b}}^{\;}_{{\bf j} \nu}$ is a $s=\frac{1}{2}$ operator
($\boldsymbol{\sigma}$ denoting Pauli matrices). The algebra satisfied
by the hard-core bosons is \cite{unveiling}: $[\bar{b}^{\;}_{{\bf
i}\sigma},\bar{b}^{\;}_{{\bf j}\sigma'} ]=0$, $[\bar{b}^{\;}_{{\bf
i}\sigma},\bar{b}^{\dagger}_{{\bf j}\sigma'} ]= \delta_{{\bf ij}}
(1-2\bar{n}_{{\bf i}\sigma}- \bar{n}_{{\bf i}\bar{\sigma}})$ (if
$\sigma=\sigma'$), or  $-\delta_{{\bf ij}} \bar{b}^{\dagger}_{{\bf
i}\sigma'}  \bar{b}^{\;}_{{\bf i}\sigma}$ (if $\sigma \neq \sigma'$).
Notice that $H$ is an extended $t$-$J$-like model of hard-core bosons
instead of constrained fermions \cite{ours1}.  These hard-core bosons
could represent three-state atoms, like the ones used in trapped
Bose-Einstein condensates, moving in an optical lattice
potential. For the sake of clarity we will only consider the AF case
$J>0$.

As explained in the introduction the first step in determining its
phase diagram consists of re-writing $H$ in a HL.
The latter is realized by $SU(3)$-spin generators in the fundamental
representation, and its mapping to the hard-core boson language can be
compactly written as \cite{unveiling}
\begin{equation}
{\cal S}({\bf j})= \begin{pmatrix} \frac{2}{3} - \bar{n}_{{\bf j}}
&\bar{b}^{\;}_{{\bf j} \uparrow} & \bar{b}^{\;}_{{\bf j} \downarrow} \\
\bar{b}^{\dagger}_{{\bf j} \uparrow}&\bar{n}_{{\bf j} \uparrow} -\frac{1}{3}&
\bar{b}^\dagger_{{\bf j} \uparrow} \bar{b}^{\;}_{{\bf j} \downarrow} \\
\bar{b}^{\dagger}_{{\bf j} \downarrow}&\bar{b}^\dagger_{{\bf j} \downarrow}
\bar{b}^{\;}_{{\bf j} \uparrow}&\bar{n}_{{\bf j} \downarrow}-\frac{1}{3}
\end{pmatrix} \ .
\label{spinsu3}
\end{equation}
The three components $s^{z}_{\bf j}=(\bar{n}_{{\bf j}\uparrow}
-\bar{n}_{{\bf j}\downarrow})/2$, $s^{+}_{\bf j}=\bar{b}^\dagger_{{\bf
j} \uparrow} \bar{b}^{\;}_{{\bf j} \downarrow}$ and $s^{-}_{\bf
j}=\bar{b}^\dagger_{{\bf j} \downarrow} \bar{b}^{\;}_{{\bf j}
\uparrow}$  generate the spin $su(2)$ subalgebra, i.e., they are the
components of the local magnetization. The five additional components
correspond to the Bose-Einstein condensate and the charge density wave
local OPs. In the HL, $H$ represents a Heisenberg-like Hamiltonian
\cite{Not5} in the presence of an external magnetic field $\mu'$
($J_{\mu \nu}=J_{\nu \mu}$)
\begin{eqnarray}
H &=& \sum_{\langle {\bf i},{\bf j} \rangle} J_{\mu \nu}
{\cal S}^{\mu \nu}({\bf i}) {\cal S}^{\nu \mu}({\bf j})
- \mu' \sum_{{\bf j}}{\cal S}^{00}({\bf j}) \ ,
\end{eqnarray}
with $J_{00}=V-J/2$, $J_{01}=J_{02}=t$,  $J_{11}=J_{12}=J_{22}=J/2$,
and $\mu'=\frac{\sf z}{3}(2V-J/2) - \bar{\mu}$. Note that through the 
mapping we transformed an interacting problem into another problem that
is quadratic in the basis of the algebra $su(3)$ in the fundamental
representation, but which is not necessarily $SU(3)$ symmetric. This 
HL furnishes the natural framework to analyze the symmetries of the
Hamiltonian $H$. There is always an $SU(2)$ spin symmetry generated by
${\cal S}^{11}-{\cal S}^{22}$, ${\cal S}^{12}$, and ${\cal S}^{21}$.
When $\mu'=0$ and $V=2t$, there are five additional generators of
symmetries related to the charge degrees of freedom. Moreover, if
$J=V=2t$ there is full $SU(3)$ symmetry. For $\mu'\ne 0$, the only
charge symmetry that remains is a $U(1)$ symmetry generated by ${\cal
S}^{00}$ (conservation of the total charge). In this way the HL,
leading to a unique OP from which all possible embedded orderings are
derived, provides a unified description of the possible thermodynamic
states of the system. Yet, it remains to establish the orderings that
survive as a result of tuning the parameters of the Hamiltonian or
external variables such as temperature and filling. 

For arbitrary values of the parameters $J/t, V/t$, we do not know a
priori how to determine exactly the phase diagram of $H$ \cite{Not3}.
The idea behind the HMFT is to perform an approximation which deals
with all possible local OPs on an equal footing with no privileged {\it
symmetry axes} and, hopefully, retains the qualitative topology of the
phase diagram. With $H$ written in the HL one immediately  realizes
that the simplest HMFT can be achieved if we re-write $H$ in terms of
$SU(3)$ Schwinger-Wigner (SW) bosons (3 flavors
$\alpha=\downarrow,0,\uparrow$) \cite{auerbach}. The mapping is
expressed as
\begin{equation}
{\cal S}({\bf j})= \begin{pmatrix} {n}_{{\bf j}0}-\frac{1}{3}&
{b}^\dagger_{{\bf j}0} {b}^{\;}_{{\bf j} \uparrow}&
{b}^\dagger_{{\bf j}0} {b}^{\;}_{{\bf j} \downarrow}\\
{b}^\dagger_{{\bf j}\uparrow}{b}^{\;}_{{\bf j}0}& 
{n}_{{\bf j} \uparrow} -\frac{1}{3}&
{b}^\dagger_{{\bf j} \uparrow} {b}^{\;}_{{\bf j} \downarrow} \\
{b}^\dagger_{{\bf j}\downarrow}{b}^{\;}_{{\bf j}0}&
{b}^\dagger_{{\bf j} \downarrow}{b}^{\;}_{{\bf j} \uparrow}&
{n}_{{\bf j} \downarrow}-\frac{1}{3}
\end{pmatrix} \ ,
\label{spinsu3b}
\end{equation}
with the SW bosons ${b}^\dagger_{{\bf j}\alpha}$ satisfying the
constraint ${n}_{{\bf j} \downarrow}+{n}_{{\bf j}0} +{n}_{{\bf j}
\uparrow}=1$. The resulting Hamiltonian ($V=2t$ with no loss of
generality) is \cite{Not5}
\begin{eqnarray}
H=\! - \! \! \sum_{\langle {\bf i},{\bf j} \rangle} (
\frac{J}{2} A^\dagger_{\bf ij}A^{\;}_{\bf ij} + t\!
\sum_{\sigma=\uparrow,\downarrow}\!
B^\dagger_{\sigma{\bf ij}}B^{\;}_{\sigma{\bf ij}} ) - \mu 
\sum_{{\bf j}}n_{{\bf j}0} \ ,
\label{hamilton}
\end{eqnarray}
where $\mu={\sf z}t -\bar{\mu}$ and the ordering operators
\begin{eqnarray}
\begin{cases}
\;\; A^\dagger_{\bf ij}= {b}^\dagger_{{\bf i} \uparrow} {
b}^\dagger_{{\bf j} \downarrow} - {b}^\dagger_{{\bf i} \downarrow} {
b}^\dagger_{{\bf j} \uparrow} \nonumber \\
B^\dagger_{\sigma{\bf ij}}= {b}^\dagger_{{\bf i} \sigma} {
b}^\dagger_{{\bf j}0} - {b}^\dagger_{{\bf i}0} {
b}^\dagger_{{\bf j} \sigma} \end{cases} \ ,
\label{orders}
\end{eqnarray}
which transform as singlets with respect to the generators of $SU(2)$
spin and charge symmetries, respectively. In other words, 
$[A^\dagger_{\bf ij}, {\cal S}^{12(21)}({\bf i})+{\cal
S}^{12(21)}({\bf j})]=0= [B^\dagger_{\uparrow(\downarrow){\bf ij}},
{\cal S}^{10(20)}({\bf i})+ {\cal S}^{10(20)}({\bf j})]$.

Since the $su(N)$ languages provide a complete set of HLs, any model
Hamiltonian can be written in a similar fashion once we identify the
appropriate HL and apply the corresponding SW mapping in the {\it
fundamental representation} (the ordering operators will, of course,
have a different meaning and algebraic expressions). The key point is
that the Hamiltonian operator in the HL becomes quadratic in the
symmetry generators of the hierarchical group. 

The idea behind any MF approximation is to disentangle interaction
terms into quadratic ones replacing some of the elementary mode
operators by their mean value.
The crux of our HMFT is that the approximation is done in the HL where
all possible local OPs are treated on an equal footing and the number
of operators replaced by their mean value is minimized since the
Hamiltonian is quadratic in the symmetry generators. In this way, the
information required is minimal. In mathematical
terms, given ${\cal O}^\dagger_{\bf ij} {\cal O}^{\;}_{\bf ij} =
\langle {\cal O}^\dagger_{\bf ij}\rangle {\cal O}^{\;}_{\bf ij}+{\cal
O}^\dagger_{\bf ij} \langle {\cal O}^{\;}_{\bf ij}\rangle - \langle
{\cal O}^\dagger_{\bf ij}\rangle \langle {\cal O}^{\;}_{\bf ij}\rangle+
({\cal O}^\dagger_{\bf ij} - \langle {\cal O}^{\dagger}_{\bf
ij}\rangle) ({\cal O}^{\;}_{\bf ij} - \langle {\cal O}^{\;}_{\bf
ij}\rangle)$, for an arbitrary bond-operator ${\cal O}^{\;}_{\bf ij}$,
the approximation amounts to neglect the latter fluctuations, i.e., 
${\cal O}^\dagger_{\bf ij} {\cal O}^{\;}_{\bf ij} \approx \langle {\cal
O}^\dagger_{\bf ij}\rangle {\cal O}^{\;}_{\bf ij}+{\cal O}^\dagger_{\bf
ij} \langle {\cal O}^{\;}_{\bf ij}\rangle - \langle {\cal
O}^\dagger_{\bf ij}\rangle \langle {\cal O}^{\;}_{\bf ij}\rangle$
\cite{dynamical}. An
important result is that all local OPs are equally treated and,
moreover, symmetries of the original Hamiltonian related to the OPs are
not broken explicitly in certain limits. 


\begin{figure}[htb]
\includegraphics[angle=0,width=8.8cm,scale=1.0]{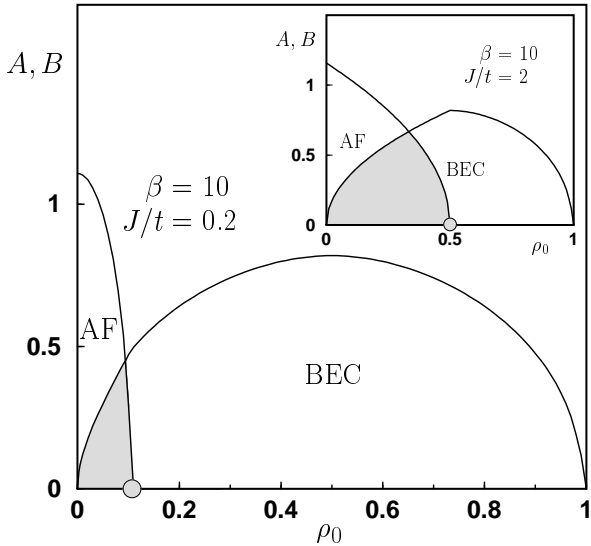}
\caption{Order parameters as a function of the density $\rho_0$ for 
different values of $J/t$ and inverse temperature $\beta=10$ (in units
of $t^{-1}$). The filled circle on the density axis indicates a quantum
critical point.}
\label{fig1}
\end{figure}
The resulting MF Hamiltonian together with the SW-boson constraint
(with Lagrange multiplier $\lambda$) $\tilde{H}=H_{MF}+\lambda
\sum_{{\bf j}} (n_{{\bf j}\downarrow} + n_{{\bf j}0} + n_{{\bf
j}\uparrow})$ reads \cite{Not5}
\begin{eqnarray}
\tilde{H}&=&\! - \! \! \sum_{\langle {\bf i},{\bf j} \rangle} 
[\frac{J A}{2} (A^\dagger_{\bf ij}+A^{\;}_{\bf ij}) + tB\!
\sum_{\sigma=\uparrow,\downarrow}\!
(B^\dagger_{\sigma{\bf ij}}+B^{\;}_{\sigma{\bf ij}}) ] \nonumber \\
&&- \mu \sum_{{\bf j}}n_{{\bf j}0} + \lambda \sum_{{\bf j}} (n_{{\bf
j}\downarrow} + n_{{\bf j}0} + n_{{\bf j}\uparrow})\nonumber \\
&=&\sum_{{\bf k}\in {\rm RBZ}} [ \Lambda_A \ b^\dagger_{{\bf k}
\uparrow} b^\dagger_{-{\bf k}+{\bf Q} \downarrow} + \Lambda_B \ (
b^\dagger_{{\bf k} \uparrow} b^\dagger_{-{\bf k}+{\bf Q}0}+\nonumber \\
&b^\dagger_{{\bf k} \downarrow}&\!\!\!\! b^\dagger_{-{\bf k}+{\bf Q}0})
+{\rm H.c.}] + (\lambda-\mu) n_{{\bf k}0}+ \!\lambda 
\!\!\!\sum_{\sigma=\uparrow,\downarrow}\!\! n_{{\bf k}\sigma}
\label{hamiltonMF}
\end{eqnarray}
where the sum of momenta ${\bf k}$ is performed over the reduced 
Brillouin zone (RBZ) with AF ordering wave vector {\bf Q} and $n_{{\bf
k}\alpha}=b^\dagger_{{\bf k}\alpha} b^{\;}_{{\bf k}\alpha}$, with
$b^\dagger_{{\bf k}\alpha}$ representing Fourier transformed modes.
$\Lambda_A=-2JA \gamma_{\bf k}$, $\Lambda_B=-4tB \gamma_{\bf k}$, with 
$\gamma_{\bf k}=\frac{1}{\sf z} \sum_{\boldsymbol{\delta}} e^{i {\bf
k}\cdot\boldsymbol{\delta}}$ ($\boldsymbol{\delta}$ are
nearest-neighbor vectors). Note that when $B=0$ in $H_{MF}$, the
$SU(2)$ spin and $U(1)$, ${\cal S}^{00}$, symmetries are conserved; the
opposite case $A=0$ preserves ${\cal S}^{10(01)}+{\cal S}^{20(02)}$
and  ${\cal S}^{11}+{\cal S}^{22}-{\cal S}^{00}$ symmetries. In 
Eq.~(\ref{hamiltonMF}) we have only considered homogeneous solutions
\cite{Not6}.

The corresponding self-consistent MF equations to solve are
\begin{eqnarray}
\begin{cases} \displaystyle 
A= \frac{8}{{\sf z}N_s} \sum_{{\bf k}\in {\rm RBZ}} \gamma_{\bf k} 
\langle b^\dagger_{{\bf k} \uparrow} b^\dagger_{-{\bf k}+{\bf
Q}\downarrow}\rangle_{MF} \ , \nonumber \\ \displaystyle
B= \frac{8}{{\sf z}N_s} \sum_{{\bf k}\in {\rm RBZ}} \gamma_{\bf k}
\langle  b^\dagger_{{\bf k} \sigma} b^\dagger_{-{\bf k}+{\bf
Q}0}\rangle_{MF} \ , \nonumber \\ \displaystyle 
1=\frac{1}{N_s}\sum_{{\bf k}\in {\rm RBZ}} \sum_{\alpha}
\langle n_{{\bf k}\alpha}\rangle_{MF}\ .
\end{cases} 
\label{scf}
\end{eqnarray}
We are thus left with a non-interacting system of SW bosons. Now we
follow Colpa \cite{colpa} and diagonalize para-unitarily the
Hamiltonian matrix $\tilde{H}$ \cite{Not1}. The application of a
homogeneous linear transformation leads to \cite{Not5}
\begin{equation}
\tilde{H} = \sum_{{\bf k}\in {\rm RBZ}} \sum_{i=0}^5 \omega_{i{\bf k}} \ 
\alpha^\dagger_{i{\bf k}} \alpha^{\;}_{i{\bf k}} \ ,
\end{equation}
where the mode energies $\omega_{i{\bf k}}$ are, at least, two-fold
degenerate \cite{Not2}. In Fig.~\ref{fig1}, we display the orders $A$
and $B$ as a function of $\rho_0=\frac{1}{N_s}\sum_{\bf j} \langle
n_{{\bf j}0}\rangle$ at very low temperature ($\beta=10$) and different
ratios of the competing interactions $J/t$ for a two-dimensional
lattice \cite{Not4}. The relation between the OPs of the original
problem, Eq.~(\ref{hamilt}), and $A$ and $B$ is given by
$\frac{1}{N_s^2} \sum_{{\bf i},{\bf j}} e^{i {\bf Q} \cdot ({\bf
r_i}-{\bf r_j})} \langle {\bar{b}}^\dagger_{{\bf i} \sigma}
{\bar{b}}^{\;}_{{\bf j} \sigma'}\rangle_{MF} \propto B^2$,
$\frac{1}{N_s^2} \sum_{{\bf i},{\bf j}} e^{i {\bf Q} \cdot ({\bf
r_i}-{\bf r_j})} \langle s^+_{\bf i}s^-_{\bf j}\rangle_{MF} \propto
A^2$, 
and justifies the labeling of the phases $\rm AF$ (antiferromagnet) and
$\rm BEC$ (Bose-Einstein condensate) in Fig.~\ref{fig1}. 
\begin{figure}[htb]
\includegraphics[angle=0,width=7.0cm,scale=1.0]{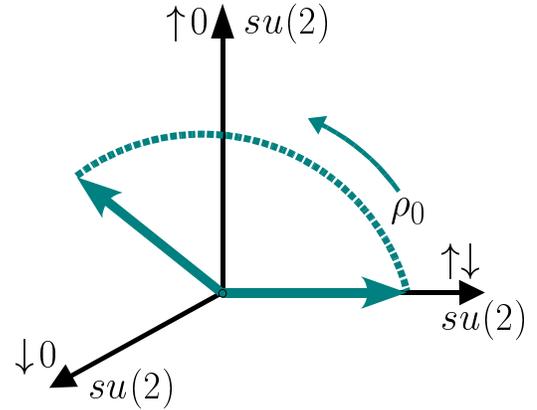}
\caption{Schematics of the order parameter change (as a function of
$\rho_0$) with the three $su(2)$ axis describing 3 different $su(2)$
subalgebras of $su(3)$. Note that the plane $\uparrow 0 \ - \!
\downarrow 0$ represents the 5 charge symmetry generators while the
$\uparrow \downarrow$ axis is associated to the remaining three 
generators of magnetism.}
\label{fig2}
\end{figure}
A way to qualitatively understand this quantum phase diagram is to look
into the OP space as displayed in Fig.~\ref{fig2}. As $\rho_0$ (or
chemical potential) varies from $0$ to $1$, the OP (depicted as
an arrow) moves in OP space. When $\rho_0=0$ the order is purely AF and
the arrow lies on the $\uparrow \downarrow$ axis. For $\rho_0\ne 0$,
the arrow has projections onto the 3 $su(2)$ axis, i.e., the AF state {\it
coexists} with a BEC state. There is a particular critical value of
$\rho_0=\rho_{0c}<1$ for which the AF ordering vanishes and the OP is
purely BEC with the arrow lying in the $\uparrow 0 \ - \! \downarrow 0$
plane. This BEC ordering persists until $\rho_0=1$, where it vanishes.


As can be inferred from our presentation, there are two complementary
aspects to studying competition and coexistence between phase orderings
in strongly coupled quantum systems. One is the direct discovery of the
{\it hidden unity} and subsequent determination of the possible phases
and their transitions, given a Hamiltonian operator modeling the
complex material of interest. This is the aspect we have described in
the present paper. The second aspect, to be discussed in a separate
publication, involves the design or engineering of new states of matter
using the inverse path of logic. Essentially, the idea consists of
tailoring effective Hamiltonians based upon a general symmetry analysis
of the possible orderings one would like to realize at zero
temperature. Tuning the parameters of these symmetry-based 
effective Hamiltonians allows one to move in parameter space along the
previously established orderings. Indeed, this strategy finds its
experimental realization in recent work done on atomic BEC systems in
optical lattices \cite{greiner}, and our approach provides a unique
theoretical guidance to achieve that goal. 
 
There are some open issues. One regards the application of the HMFT
approach to study fermionic problems, for example, a Hamiltonian like
Eq.~(\ref{hamilt}) but where the operators $\bar{b}^\dagger_{{\bf i}
\sigma}$ for different modes on a lattice ${\bf i}$ anticommute. The
method could certainly be used, however, fermions do introduce a
non-local gauge potential \cite{ours1} leading to an effective
dynamical frustration which is difficult to handle in a controlled
manner. Another issue concerns the application of the HMFT method when
longer-range interactions are involved. There is already evidence from
work on the $J_1$-$J_2$ $SU(2)$ Heisenberg model \cite{trumper} that
our HMFT will work in those cases as far as homogeneous phases are
concerned. Actually, the SW MF theory introduced by Arovas and Auerbach
\cite{arovas} in the fundamental representation is a
particular case of our general HMFT. Finally, our methodology exhausts
all broken symmetry instances but it is still quite possible to have
purely topological quantum orders and their corresponding phase
transitions which cannot be described by broken symmetries and 
associated OPs \cite{wen} and, thus, are not included in our framework.

Summarizing, we developed a theoretical framework and a calculational 
scheme to study coexistence and competition of thermodynamic phases in
strongly correlated matter. In our method (given a Hamiltonian modeling
the physical system) the order parameters are not guessed but
rigorously determined from group theoretical considerations as symmetry
generators of a hierarchical language. In this way, the Hamiltonian
operator (which does not necessarily have the full symmetry of the
hierarchical group) is expressed in terms of symmetry generators. Then,
in a non-phenomenological approach dubbed {\it hierarchical mean-field
theory}, we approximated the dynamics (and thermodynamics) treating all
possible local order parameters on an equal footing, i.e., without
preferred symmetry axis. One could say that this procedure follows the
guiding principles of {\it maximum symmetry} and {\it minimum
information}. This allowed us to obtain in a simple manner the phase
diagram of a model problem exhibiting coexistence and competition
between antiferromagnetism and superfluidity. Combined with an analysis
of the fluctuations (to analyze the stability of the mean-field) one
now has a simple machinery to design phase diagrams.



This work was sponsored by the US DOE. 
We thank J.E. Gubernatis for a careful reading of the manuscript.


\begin{references}
\bibitem{Not0}
Formally, a language is defined by a set of operators each acting
irreducibly on a local Hilbert space of dimension $D$. The language
does not determine the quantum dynamics of the system (which is
determined by the Hamiltonian) but the dynamics can be expressed in
different languages. See Ref.~\cite{rpmbt}.
\bibitem{Not5}
In the present manuscript the irrelevant constant terms in the
transformed Hamiltonian operators are always omitted.
\bibitem{Not3} 
We know that for $J=V=2t=\frac{2\bar{\mu}}{\sf z}<0$, we can find the
exact ground and lowest energy states of $H$ \cite{unveiling}.
\bibitem{dynamical}
One can add temporal fluctuations through a similar type of dynamical
mean-field as described in Ref.~\cite{dyn} but in the HL.
\bibitem{Not1}
Find a matrix $T^{-1}$ such that $(T^\dagger)^{-1} \tilde{H} T^{-1}
= {\rm diagonal}$,  and $T$ satisfies the para-unitarity condition
$T^\dagger \Punitary T =  T \Punitary T^\dagger = \Punitary$ with
$\Punitary=\begin{pmatrix} \one& \Zero \\ \Zero & -\one
\end{pmatrix}$. The necessary and sufficient condition for an Hermitian
matrix to be para-unitarily diagonalized (with all diagonal elements
positive) is that the matrix be positive definite.
\end{references}
\end{document}